\documentclass[12pt,a4paper]{article}
\usepackage{amssymb}
\usepackage{amsmath}
\usepackage{float}
\usepackage{empheq}
\begin{document}

\begin{center}
{\Large \bf Some problems with reproducing \\ the Standard
Model fields and interactions in\\\vspace{0.25cm}
five-dimensional warped brane world models}

\vspace{4mm}

Mikhail N.~Smolyakov, Igor P. Volobuev\\
\vspace{0.5cm} Skobeltsyn Institute of Nuclear Physics, Lomonosov
Moscow State University,
\\ 119991, Moscow, Russia
\end{center}

\begin{abstract}
In the present paper we examine, from the purely theoretical point
of view and in a model-independent way, the case, when matter,
gauge and Higgs fields are allowed to propagate in the bulk of
five-dimensional brane world models with compact extra dimension,
and the Standard Model fields and their interactions are supposed
to be reproduced by the corresponding zero Kaluza-Klein modes. An
unexpected result is that in order to avoid possible pathological
behavior in the fermion sector, it is necessary to impose
constraints on the fermion field Lagrangian. In the case when the
fermion zero modes are supposed to be localized at one of the
branes, these constraints imply an additional relation between the
vacuum profile of the Higgs field and the form of the background
metric. Moreover, this relation between the vacuum profile of the
Higgs field and the form of the background metric results in the
{\em exact} reproduction of the gauge boson and fermion sectors of
the Standard Model by the corresponding zero mode four-dimensional
effective theory in all the physically relevant cases, allowed by
the absence of pathologies. Meanwhile, deviations from
these conditions can lead either back to pathological behavior in
the fermion sector or to a variance between the resulting zero
mode four-dimensional effective theory and the Standard Model,
which, depending on the model at hand, may, in principle, result
in constraints putting the theory out of the reach of the present
day experiments.
\end{abstract}

\section{Introduction}
Models with extra dimensions have been attracting a great interest
during the last fifteen years. There were many attempts to solve
various theoretical problems  with the help of extra dimensions. A
wide branch of multidimensional models is that of brane world
models, which were proposed  in their modern form in
\cite{ArkaniHamed:1998rs,Randall:1999ee}. Although some
theoretical problems (such as, for example, the hierarchy problem
of gravitational interaction) were successfully solved within the
framework of brane world models, realistic theories must also
describe all the physical aspects of our four-dimensional world.
In particular, they must correctly reproduce the interactions of
the Standard Model (SM) particles that have already been tested
experimentally.

In the original formulation of brane world models the SM fields were supposed
to be located on a brane (in the Randall-Sundrum model \cite{Randall:1999ee},
on the TeV brane). Later the idea of brane worlds was
joined with the idea that all the fields can propagate in extra dimensions
\cite{Appelquist:2000nn} thus giving rise to the
theories with universal extra dimensions, where the matter, gauge and Higgs fields
are allowed to propagate in the bulk of five-dimensional brane
world models with compact extra dimension. In this case all these
fields posses towers of Kaluza-Klein excitations, their zero modes being the
SM fields. There exist many papers describing how the SM can be embedded this way into
multidimensional brane worlds and what new effects can be produced
in such theories.

However, there are some finer points that have been missed in the
previous studies. Below we will discuss them in detail from a
purely theoretical point of view and in a mathematically
consistent way. In particular, we argue that it is impossible to
{\it exactly} reproduce by the zero Kaluza-Klein modes the
electroweak gauge boson sector of the SM in the effective
four-dimensional theory, unless the vacuum profile of the Higgs
field in the extra dimension behaves like the square root of the
inverse warp factor. The same vacuum profile of the Higgs field
(together with extra constraints on the parameters of the
five-dimensional fermion field Lagrangian) is necessary for
reproducing the fermion sector of the SM. An unexpected result is
that deviations from these conditions may lead either to
pathologies in the fermion sector when expanding in Kaluza-Klein
modes (for example, in the case admitting consistent localization
of the fermion zero mode at one of the branes) or to an additional
modification of the couplings of fermions to gauge bosons in the
zero mode sector, which may lead, in principle, to severe
restrictions on the value of the five-dimensional energy scale
(like the constraints coming from the zero mode gauge
boson sector \cite{Csaki:2002gy,Burdman:2002gr}). Meanwhile, the
demand for the extra restriction on the vacuum profile of the
Higgs field, mentioned above, leads to some difficulties, such as
the necessity for an extra fine-tuning. This problem needs to be
addressed, at least for better understanding the structure of
brane world models.

The paper is organized as follows. In Section~2 we consider an
illustrative example of bulk gauge fields interacting with the
bulk Higgs scalar field and show which conditions should be
fulfilled in order the gauge fields exactly reproduce the electroweak gauge
sector of the SM by the lowest Kaluza-Klein modes. In Section~3
fermions are examined in the same way. In Section~4 we consider
interactions between fermions and gauge bosons in a simple theory,
where pathologies in the fermion sector are absent, whereas no
extra conditions on the Higgs field are imposed. The obtained
results are discussed in the last section.

\section{Gauge fields}
Let us take a five-dimensional space-time with the coordinates
$x^{M}=\{x^{\mu},z\}$, $M=0,1,2,3,5$. The compact extra dimension
is supposed to form the orbifold $ S^1/Z_2$, which can be
represented as the circle with the coordinate $-L\le z\le L$ and
the points $-z$ and $z$ identified. In what follows, we will use
the notation $x$ for the coordinates $x^{\mu}$. We consider the
following standard form of the background metric, which is often
used in brane world models:
\begin{equation}\label{backgmetric}
ds^2=e^{2\sigma(z)}\eta_{\mu\nu}dx^{\mu}dx^{\nu}-dz^2.
\end{equation}
This metric is assumed to correspond to a regular brane
world model, i.e. it is a solution to equations of motion for
five-dimensional gravity, two branes with tension and, for
example, a stabilizing bulk scalar field. We do not specify the
explicit form of the  solution for $\sigma(z)$.

We start with the gauge fields and choose the following action of
an $SU(2)\times U(1)$ gauge invariant model in this background:
\begin{equation}\label{action}
S=\int d^{4}xdz\sqrt{g}\left(-\frac{\xi^2}{4}F^{a,MN}F^{a}_{MN}-\frac{\xi^2}{4}B^{MN}B_{MN}+g^{MN}\left(D_{M}H\right)^{\dag}D_{N}H-V(H^{\dag}H)\right),
\end{equation}
where
\begin{eqnarray}\label{F1}
&&F^{a}_{MN}=\partial_{M}A^{a}_{N}-\partial_{N}A^{a}_{M}+g\epsilon^{abc}A^{b}_{M}A^{c}_{N},\\ \label{F1a}
&&B_{MN}=\partial_{M}B_{N}-\partial_{N}B_{M},\\ \label{F2}
&&D_{M}H=\left(\partial_{M}-ig\frac{\tau^{a}}{2}A^{a}_{M}-i\frac{g'}{2}B_{M}\right)H
\end{eqnarray}
and the fields satisfy the orbifold symmetry conditions $A^{a}_{\mu}(x,-z)=A^{a}_{\mu}(x,z)$, $A^{a}_{5}(x,-z)=-A^{a}_{5}(x,z)$, $B_{\mu}(x,-z)=B_{\mu}(x,z)$, $B_{5}(x,-z)=-B_{5}(x,z)$, $H(x,-z)=H(x,z)$. Here $\xi=\frac{1}{\sqrt{2L}}$ is a constant, which is introduced for convenience and chosen
so that the dimension of the bulk gauge fields is mass. The scalar field potential can include brane-localized terms of the form $\lambda_{1}(H^{\dag}H)\delta(z)$ and
$\lambda_{2}(H^{\dag}H)\delta(z-L)$. It is easy to see that action
(\ref{action}),  which has a rather standard form, resembles the
bosonic sector of the electroweak part of the ordinary
four-dimensional SM.

This action gives rise to the equations of motion for the gauge
and the Higgs fields that look like
\begin{eqnarray}\label{eq_gauge}
&&\nabla_N F^{a,MN}+g\epsilon^{abc}A^{b}_{N}F^{c,MN} + i\frac{g}{\xi^2}\left(\left(D^{M}H\right)^{\dag}\frac{\tau^{a}}{2}H  - H^{\dag}\frac{\tau^{a}}{2}D^{M}H\right) = 0,\\ \label{eq_gauge2}
&&\nabla_N B^{MN} + i\frac{g'}{2{\xi^2}}\left(\left(D^{M}H\right)^{\dag}H  - H^{\dag}D^{M}H\right) = 0,\\ \label{eq_gauge3}
&&\nabla^{M} D_{M}H -
g^{MN}\left(ig\frac{\tau^{a}}{2}A^{a}_{M}+i\frac{g'}{2}B_{M}\right)
D_{N}H + \frac{d V}{d (H^{\dag}H)}H = 0,
\end{eqnarray}
 $\nabla^M$ denoting the covariant derivative with respect to
metric (\ref{backgmetric}).

Let us consider the vacuum solution for these fields. The vacuum solution, breaking the gauge group $SU(2)\times U(1)$ to
$U(1)_{em}$, leaving the Poincare invariance in four-dimensional space-time intact and satisfying equations (\ref{eq_gauge}), (\ref{eq_gauge2}), can be taken in the form
\begin{equation}\label{vev}
A^{a}_{M}\equiv 0,\qquad B_{M}\equiv 0,\qquad
H_{0}\equiv\left(0\atop \frac{v(z)}{\sqrt{2}}\right),
\end{equation}
where $v(z)$ is a real function. It is not difficult to understand
that in the general case the vacuum solution for the Higgs field
$v(z)$ may depend on the coordinate of the extra dimension. Of
course, the scalar field potential must provide for such a
solution to equation (\ref{eq_gauge3}). At this point we do not specify the explicit form of
$v(z)$.

Now let us turn to examining the excitations in the model at hand.
Below we will be interested in the behavior of only the
four-vector components of the five-dimensional gauge fields, whose
zero modes must play the role of the SM gauge bosons. For this
reason, from here on we retain only these components of the gauge
fields and drop the components $A^{a}_{5}$, $B_{5}$ of the vector fields and
the fluctuations of the Higgs field. From action (\ref{action}) it is easy to
get the following effective action for the four-vector components of
the five-dimensional gauge fields:
\begin{eqnarray}\label{effact}
S_{eff}=\int d^{4}xdz\left(-\frac{\xi^2}{4}\eta^{\mu\nu}\eta^{\alpha\beta}F^{a}_{\mu\alpha}F^{a}_{\nu\beta}+
e^{2\sigma}\frac{\xi^2}{2}\eta^{\mu\nu}\partial_{5}A^{a}_{\mu}\partial_{5}A^{a}_{\nu}
-\frac{\xi^2}{4}\eta^{\mu\nu}\eta^{\alpha\beta}B_{\mu\alpha}B_{\nu\beta}\right.\\ \nonumber
\left.+e^{2\sigma}\frac{\xi^2}{2}\eta^{\mu\nu}\partial_{5}B_{\mu}\partial_{5}B_{\nu}+
e^{2\sigma}\eta^{\mu\nu}H^{\dag}_{0}\left(g\frac{\tau^{a}}{2}A^{a}_{\mu}+
\frac{g'}{2}B_{\mu}\right)\left(g\frac{\tau^{a}}{2}A^{a}_{\nu}+\frac{g'}{2}B_{\nu}\right)H_{0}\right),
\end{eqnarray}
where we have also dropped the terms containing only the vacuum configuration of the Higgs field.

Now we are ready to perform the Kaluza-Klein mode decomposition.
First, using the standard redefinition
\begin{eqnarray}\label{gaugephys}
Z_{\mu}=\frac{1}{\sqrt{g^2+{g'}^2}}\left(gA^{3}_{\mu}-g'B_{\mu}\right),\,\, A_{\mu}=\frac{1}{\sqrt{g^2+{g'}^2}}\left(g B_{\mu}+g'A^{3}_{\mu}\right),\,\,
W^{\pm}_{\mu}=\frac{1}{\sqrt{2}}\left(A^{1}_{\mu}\mp iA^{2}_{\mu}\right),
\end{eqnarray}
we can pass to the physical degrees of freedom of the theory.
Next, let us consider only the quadratic part of effective action
(\ref{effact}) in terms of these new fields. It takes the form
\begin{eqnarray}\label{effact2}
S_{eff}=\int d^{4}xdz\left(-\frac{\xi^2}{2}\eta^{\mu\nu}\eta^{\alpha\beta}W^{+}_{\mu\alpha}W^{-}_{\nu\beta}
-\frac{\xi^2}{4}\eta^{\mu\nu}\eta^{\alpha\beta}F_{\mu\alpha}F_{\nu\beta}-\frac{\xi^2}{4}\eta^{\mu\nu}\eta^{\alpha\beta}Z_{\mu\alpha}Z_{\nu\beta}\right.\\ \nonumber
 \left.
+e^{2\sigma}\xi^2\eta^{\mu\nu}\partial_{5}W^{+}_{\mu}\partial_{5}W^{-}_{\nu}+e^{2\sigma}\frac{\xi^2}{2}\eta^{\mu\nu}\partial_{5}A_{\mu}\partial_{5}A_{\nu}
+e^{2\sigma}\frac{\xi^2}{2}\eta^{\mu\nu}\partial_{5}Z_{\mu}\partial_{5}Z_{\nu}\right.\\ \nonumber
 \left.+e^{2\sigma}v^{2}(z)\frac{g^{2}}{4}\eta^{\mu\nu}W^{+}_{\mu}W^{-}_{\nu}+e^{2\sigma}v^{2}(z)\frac{g^2+{g'}^{2}}{8}\eta^{\mu\nu}Z_{\mu}Z_{\nu}\right),
\end{eqnarray}
where $W^{\pm}_{\mu\nu}=\partial_{\mu}W^{\pm}_{\nu}-\partial_{\nu}W^{\pm}_{\mu}$, $F_{\mu\nu}=\partial_{\mu}A_{\nu}-\partial_{\nu}A_{\mu}$, $Z_{\mu\nu}=\partial_{\mu}Z_{\nu}-\partial_{\nu}Z_{\mu}$. The equations for the wave functions and the masses of the Kaluza-Klein modes are
\begin{eqnarray}\label{Wwf}
-m_{W,n}^{2}f_{W,n}-\partial_{5}(e^{2\sigma}\partial_{5}f_{W,n})+\frac{g^2}{4\xi^2}e^{2\sigma}v^2(z)f_{W,n}=0,\\ \label{Zwf}
-m_{Z,n}^{2}f_{Z,n}-\partial_{5}(e^{2\sigma}\partial_{5}f_{Z,n})+\frac{g^2+{g'}^{2}}{4\xi^2}e^{2\sigma}v^2(z)f_{Z,n}=0,\\ \label{Awf}
-m_{A,n}^{2}f_{A,n}-\partial_{5}(e^{2\sigma}\partial_{5}f_{A,n})=0.
\end{eqnarray}
where the subscript $n$ denotes the number of the corresponding
Kaluza-Klein mode.  As usual, the lowest (zero)
Kaluza-Klein modes of the fields are supposed to correspond to the
four-dimensional SM particles. So, below we will focus only on
the zero modes.

It follows from (\ref{Awf}) that the solution for the lowest mode
of the field $A_{\mu}$ (the photon) is $m_{A,0}=0$ and
$f_{A,0}(z)\equiv\textrm{const}$, i.e. its wave function does not
depend on the coordinate of the extra dimension. This is an
important result, which provides the universality of the
electromagnetic charge \cite{Rubakov}. But, as one sees from
(\ref{Wwf}) and (\ref{Zwf}),  in the general case it is not so for
the zero modes of the fields $W_{\mu}$ and $Z_{\mu}$, which
correspond to the SM massive gauge bosons. The latter has the
following well-known consequences. Indeed, in the SM the
self-coupling of massive gauge bosons comes from the term
$F^{a,\mu\nu}F^{a}_{\mu\nu}$ and the corresponding coupling
constants are defined only by the structure of the gauge group. In
the five-dimensional case under consideration the
self-coupling terms also come from the same term of
(\ref{effact}), but now the corresponding coupling constants are
also defined by the overlap integrals over the coordinate $z$,
which include the wave functions $f_{W,0}(z)$ and $f_{Z,0}(z)$.
The only case, when the zero mode sector of the model automatically completely
coincides with the electroweak gauge boson sector of the SM,
is the one, where the wave functions $f_{W,0}(z)$ and $f_{Z,0}(z)$
do not depend on the coordinate of the extra dimension. In this
case the self-coupling constants of the massive gauge bosons are
defined in terms of the constants $g$ and $g'$ exactly in the same
way as in the ordinary SM. The independence of the wave functions
$f_{W,0}(z)$ and $f_{Z,0}(z)$ on the coordinate of the extra
dimension can be achieved only when
$e^{2\sigma}v^2(z)\equiv\textrm{const}$, i.e., when
\begin{equation}\label{higgsprof}
v(z)\equiv\xi\tilde v e^{-\sigma},
\end{equation}
where $\tilde v$ is a constant of dimension $M$. For the choice
(\ref{higgsprof}), the masses of the zero mode gauge bosons are given by
\begin{equation}\label{gmasses}
m_{W,0}=\frac{g \tilde v}{2},\qquad
m_{Z,0}=\frac{\sqrt{g^2+{g'}^{2}}\tilde v}{2}.
\end{equation}
Thus, in the case under consideration  $\tilde v$ must
coincide with the Higgs field vacuum expectation value of the SM.

Of course, the results presented above are rather trivial.
Moreover, it is well known that in the general case a modification
of the shapes of the zero mode gauge boson wave functions has an
influence on the electroweak observables, this problem was
discussed in detail in \cite{Csaki:2002gy,Burdman:2002gr}. It is
shown in these papers that, for example, in the case of the
Randall-Sundrum model \cite{Randall:1999ee} such a modification
leads to  restrictions on the value of the five-dimensional energy
scale, which put the theory out of the reach of the present day
experiments. For the choice (\ref{higgsprof}) the lowest mode
four-dimensional effective theory exactly reproduces the
electroweak gauge boson sector of the SM, thus imposing no
restrictions on the value of the five-dimensional energy scale.
Meanwhile, one can imagine that there exists a profile for the
Higgs vacuum solution, which differs from (\ref{higgsprof}) but
provides somehow the necessary values of the zero mode gauge boson
masses and self-coupling constants with a good accuracy.
Unfortunately, the situation becomes more involved, when one comes
to fermions.

\section{Fermions}
It is well known that, since there is no chirality in
five-dimensional  space-time, in order to obtain a nonzero
mass term for the zero Kaluza-Klein fermion mode via the Higgs
mechanism it is necessary to take two five-dimensional spinor
fields (see, for example, \cite{DRT,Macesanu,CG}) satisfying the orbifold
symmetry conditions
\begin{eqnarray}\label{sym1}
\Psi_{1}(x,-z)=\gamma^{5}\Psi_{1}(x,z),\\ \label{sym2}
\Psi_{2}(x,-z)=-\gamma^{5}\Psi_{2}(x,z).
\end{eqnarray}
Thus, as a simple example, we consider a  model with  the action of
the most general form
\begin{eqnarray}\label{faction}
S=\int d^{4}xdz\sqrt{g}\left(E_{N}^{M}i\bar\Psi_{1}\Gamma^{N}\nabla_{M}\Psi_{1}+E_{N}^{M}i\bar\Psi_{2}\Gamma^{N}\nabla_{M}\Psi_{2}\right.\\
\nonumber\left.-F_{1}(z)\bar\Psi_{1}\Psi_{1}-
F_{2}(z)\bar\Psi_{2}\Psi_{2}-G(z)\left(\bar\Psi_{2}\Psi_{1}+\bar\Psi_{1}\Psi_{2}\right)\right),
\end{eqnarray}
where $M,N=0,1,2,3,5$, $\Gamma^{\mu}=\gamma^{\mu}$,
$\Gamma^{5}=i\gamma^{5}$, $\nabla_{M}$ is the covariant derivative
containing  the spin connection, $E_{N}^{M}$ is the vielbein,
$F_{1,2}(z)$ and $G(z)$ are some functions satisfying the symmetry
conditions $F_{1,2}(-z)=-F_{1,2}(z)$ and $G(-z)=G(z)$. For the
case of metric (\ref{backgmetric}) action (\ref{faction}) can be
rewritten in the form (see, for example, \cite{CG,fRS} for the
explicit form of the vielbein and spin connections)
\begin{eqnarray}\label{feffact}
S=\int d^{4}xdze^{4\sigma}\left(e^{-\sigma}i\bar\Psi_{1}\gamma^{\mu}\partial_{\mu}\Psi_{1}-\bar\Psi_{1}\gamma^{5}\left(\partial_{5}+2\sigma'\right)\Psi_{1}
-F_{1}(z)\bar\Psi_{1}\Psi_{1}\right.\\ \nonumber\left.
+e^{-\sigma}i\bar\Psi_{2}\gamma^{\mu}\partial_{\mu}\Psi_{2}-\bar\Psi_{2}\gamma^{5}\left(\partial_{5}+2\sigma'\right)\Psi_{2}
-F_{2}(z)\bar\Psi_{2}\Psi_{2}-G(z)\left(\bar\Psi_{2}\Psi_{1}+\bar\Psi_{1}\Psi_{2}\right)\right),
\end{eqnarray}
where $'=\partial_{5}$. The equations of motion,
following from this action, take the form
\begin{eqnarray}\label{feqs1}
e^{-\sigma}i\gamma^{\mu}\partial_{\mu}\Psi_{1}-\gamma^{5}\left(\partial_{5}+2\sigma'\right)\Psi_{1}
-F_{1}(z)\Psi_{1}-G(z)\Psi_{2}=0,\\ \label{feqs2}
e^{-\sigma}i\gamma^{\mu}\partial_{\mu}\Psi_{2}-\gamma^{5}\left(\partial_{5}+2\sigma'\right)\Psi_{2}
-F_{2}(z)\Psi_{2}-G(z)\Psi_{1}=0.
\end{eqnarray}

Suppose that $G(z)\equiv 0$. In this case there always exists the solution
\begin{equation}\label{masslessf}
\Psi_{1}=C_{f}\exp\left[-\int\limits_{0}^{z}F_{1}(y)dy-2\sigma(z)\right]\psi_{L}(x),\qquad i\gamma^{\mu}\partial_{\mu}\psi_{L}=0,\qquad \gamma^{5}\psi_{L}=\psi_{L},
\end{equation}
where $C_{f}$ is a normalization constant, describing a massless
four-dimensional fermion. An analogous solution exists for the
field $\Psi_{2}$ (but with a right-handed four-dimensional
fermion). The latter clearly indicates that the existence of only
one five-dimensional fermion is not enough to provide a massive
four-dimensional lowest mode. This also indicates that it is the
term with $G(z)\not\equiv 0$ that is responsible for the
generation of the masses of the zero mode fermions. Thus, the
function $G(z)$ should be somehow connected with the
five-dimensional Higgs field. It is natural to take this function
as
\begin{equation}\label{Gofz}
G(z)\equiv hv(z),
\end{equation}
where $h$ is a coupling constant of dimension $M^{-\frac{1}{2}}$.
Such a construction may arise, when one considers the standard
Higgs mechanism in the bulk after the spontaneous symmetry breaking, leading
to (\ref{vev}), whereas the zero modes of the fields $\Psi_{1}$, $\Psi_{2}$ are supposed
to represent a massive lepton (for example, the electron).

It should be noted that the ``localizing'' functions $F_{1}(z)$
and $F_{2}(z)$ are not connected with the Higgs field in the
general case. Meanwhile, the corresponding terms are not forbidden
and, according to (\ref{masslessf}), they are responsible for the
localization of the lowest fermion Kaluza-Klein modes. In fact,
the form of the terms with the functions $F_{1}(z)$ and $F_{2}(z)$
in (\ref{faction}) is the only one suitable for the
localization of the fermion zero modes in a consistent
field-theoretical manner, though the origin of the localizing
functions can be different. In the context of multidimensional
models, such a mechanism was proposed in \cite{RS} for the theory
with one infinite extra dimension,  in which the localizing
function was just a profile of the topological soliton modeling a
domain wall. An analogous mechanism (but with another form of the
localizing function) was used in \cite{ArkaniHamed:1999dc}, where
fermion fields were also supposed to be confined to a thick domain
wall in flat extra dimensions, but different fermions were
localized at different points of the wall. In five-dimensional
brane world models the localizing functions are usually  not
explicitly connected with something like a domain wall, very often
the corresponding terms have the form similar to the standard
fermion mass term, but with an antisymmetric ``mass'' according to
the orbifold symmetry, i.e., $F(z)=C\,\textrm{sign}(z)$. The
value of the constant $C$ defines at which brane the fermion zero mode
is localized and what is the width of its wave function
\cite{Grossman:1999ra,Gherghetta:2000qt,Huber:2000ie}.

Now let us recall the ordinary four-dimensional free spinor field
satisfying the Dirac equation. It is well known that each
component of this field satisfies the Klein-Gordon equation, which
is a second-order differential equation. Of course, all the components of the spinor field are not independent --- with the help of the initial Dirac equation one can restore, for example, the two-component spinor $\psi_{R}$ using a solution for the two-component spinor $\psi_{L}$, where the components of $\psi_{L}$ are supposed to be independent and to satisfy the Klein-Gordon equation. The Klein-Gordon equation is known to have no pathologies, so one can be sure that the whole theory is consistent. The five-dimensional fields $\Psi_{1}$ and $\Psi_{2}$ satisfying equations
(\ref{feqs1}), (\ref{feqs2}) should be considered as free fields
as well, because they are coupled only to the vacuum
configurations of the Higgs and gravity fields. Therefore, one
expects that in a consistent theory {\em each component} of the
five-dimensional spinor fields $\Psi_{1}$ and $\Psi_{2}$ (or at
least of their linear combinations) also satisfies a
five-dimensional second-order differential equation, which
contains derivatives in the four-dimensional coordinates only in
the form $\Box=\eta^{\mu\nu}\partial_{\mu}\partial_{\nu}$,
otherwise one may expect the appearance of various pathologies
when expanding in Kaluza-Klein modes (which is, in fact, the first
step in examining the four-dimensional effective theory), an
example of such a pathological behavior will be presented below.
The latter is not good taking into account the fact that the
components of the five-dimensional spinors make up
four-dimensional fermion fields (just like how $\psi_{L}$ and $\psi_{R}$ make up a four-dimensional four-component spinor field), see, for example, \cite{Macesanu,Smolyakov:2011hv}. So, let us try to obtain the
corresponding second-order differential equations. From
(\ref{feqs1}) and (\ref{feqs2}) it is not difficult to obtain:
\begin{eqnarray}\label{KGfeqs1}
-\Box\Psi_{1}+e^{\sigma}(\partial_{5}+2\sigma')e^{\sigma}(\partial_{5}+2\sigma')\Psi_{1}+
e^{\sigma}\partial_{5}(e^{\sigma}F_{1}(z))\gamma^{5}\Psi_{1}
-e^{2\sigma}(F_{1}^{2}(z)+h^2v^{2}(z))\Psi_{1}\\ \nonumber +
he^{\sigma}\partial_{5}(e^{\sigma}v(z))\gamma^{5}\Psi_{2}-he^{2\sigma}v(z)\left(F_{1}(z)+F_{2}(z)\right)\Psi_{2}=0,\\
\label{KGfeqs2}
-\Box\Psi_{2}+e^{\sigma}(\partial_{5}+2\sigma')e^{\sigma}(\partial_{5}+2\sigma')\Psi_{2}+e^{\sigma}\partial_{5}(e^{\sigma}F_{2}(z))\gamma^{5}\Psi_{2}
-e^{2\sigma}(F_{2}^{2}(z)+h^2v^{2}(z))\Psi_{2}\\ \nonumber
+he^{\sigma}\partial_{5}(e^{\sigma}v(z))\gamma^{5}\Psi_{1}-he^{2\sigma}v(z)\left(F_{1}(z)+F_{2}(z)\right)\Psi_{1}=0.
\end{eqnarray}
From (\ref{KGfeqs1}) and (\ref{KGfeqs2}) one can see that formally
the equations for the components of the fields $\Psi_{1}$ and
$\Psi_{2}$ do not decouple. It turns out that in the general case
we can not obtain second-order differential equations
for each component of the fields $\Psi_{1}$ and $\Psi_{2}$ (or of
their linear combinations) separately, as it happens in the
ordinary four-dimensional theory, except several special cases.
The first obvious exception is when the following conditions
fulfill:
\begin{eqnarray}\label{cond1}
F_{1}(z)\equiv-F_{2}(z),\\ \label{cond2}
\partial_{5}(e^{\sigma}v(z))\equiv 0.
\end{eqnarray}
The second condition completely coincides with (\ref{higgsprof}).
Introducing the dimensionless coupling constant $\tilde h = h
\xi$ and taking into account   (\ref{cond1}) and (\ref{cond2}) we
can rewrite  equations (\ref{KGfeqs1}) and (\ref{KGfeqs2}) as
\begin{eqnarray}\label{KGfeqs1a}
-\Box\Psi_{1}+e^{\sigma}(\partial_{5}+2\sigma')e^{\sigma}(\partial_{5}+2\sigma')\Psi_{1}+e^{\sigma}\partial_{5}(e^{\sigma}F)\gamma^{5}\Psi_{1}
-(e^{2\sigma}F^{2}+{\tilde h}^2{\tilde v}^{2})\Psi_{1}=0,\\
\label{KGfeqs2a}
-\Box\Psi_{2}+e^{\sigma}(\partial_{5}+2\sigma')e^{\sigma}(\partial_{5}+2\sigma')\Psi_{2}-e^{\sigma}\partial_{5}(e^{\sigma}F)\gamma^{5}\Psi_{2}
-(e^{2\sigma}F^{2}+{\tilde h}^2{\tilde v}^{2})\Psi_{2}=0,
\end{eqnarray}
where $F(z)\equiv F_{1}(z)\equiv-F_{2}(z)$, which indeed lead to the second-order differential equations for
each component of the fields $\Psi_{1}$ and $\Psi_{2}$
separately.

The solution to these equations, corresponding to the zero mode,
has the form (it can also  be easily obtained from (\ref{feqs1}),
(\ref{feqs2}))
\begin{eqnarray}\label{massf1}
\Psi_{1}=C_{f}\exp\left[-\int\limits_{0}^{z}F(y)dy-2\sigma(z)\right]\psi_{L}(x),\quad
i\gamma^{\mu}\partial_{\mu}\psi_{L}-\tilde h\tilde v\psi_{R}=0,&&
\gamma^{5}\psi_{L}=\psi_{L},\\ \label{massf2}
\Psi_{2}=C_{f}\exp\left[-\int\limits_{0}^{z}F(y)dy-2\sigma(z)\right]\psi_{R}(x),\quad
i\gamma^{\mu}\partial_{\mu}\psi_{R}- \tilde h\tilde v\psi_{L}=0,&&
\gamma^{5}\psi_{R}=-\psi_{R},
\end{eqnarray}
where again $C_{f}$ is a normalization constant. This solution
indeed corresponds to the lowest mode, see Appendix~A for details.
It is clear that the fields $\psi_{L}$ and $\psi_{R}$ are
localized in the vicinity of the same point in the extra
dimension. Taken together they make up a four-dimensional Dirac
fermion with mass $\tilde h\tilde v$. As for the physical degrees of freedom corresponding to higher Kaluza-Klein fermion modes, for the case of equations (\ref{KGfeqs1a}), (\ref{KGfeqs2a}) they can be examined exactly in the same way as it was made in \cite{Smolyakov:2011hv} for the model with infinite extra dimension.

It should be also mentioned that the fermion action exactly of
form (\ref{feffact}) with conditions (\ref{cond1}) and (\ref{cond2}) (but in other notations) was considered in \cite{CG}
for examining discrete symmetries in brane world models.

It is interesting to note that if the localizing functions $F(z)$
have one and the same form for all fermion fields in the theory
(leptons, quarks), then the wave functions of the zero modes also
have the same form for different fermions regardless of the
four-dimensional mass of the mode (see (\ref{massf1}),
(\ref{massf2})). In this case the coupling constants of fermions
to gauge bosons in the zero  mode sector appear to be exactly the
same as in the SM. This happens because the wave functions of all
the zero mode gauge bosons do not depend on the coordinate of the
extra dimension if relation (\ref{cond2}) holds (see Section~2) and all the
corresponding vertices (even those containing two different
four-dimensional fermions) in fact contain the integral of the
same fermion wave function squared, and this integral is equal to
unity due to the normalization conditions.

The question arises, whether there are other exceptions in
equations (\ref{KGfeqs1}), (\ref{KGfeqs2}), leading to
second-order differential equations of motion for any form of
$v(z)$? We found another simple exception for the general case (in
principle, there are more exceptions, but they seem to be very
unnatural, see Appendix~B for a detailed discussion of the
decoupling of equations (\ref{KGfeqs1}), (\ref{KGfeqs2})), which
is, in fact, rather obvious and follows even from the form of
equations (\ref{KGfeqs1}), (\ref{KGfeqs2}). Namely, if the
relation
\begin{equation}\label{equivcond}
F_{1}(z)\equiv F_{2}(z)
\end{equation}
is fulfilled, then one can simply add and subtract equations (\ref{KGfeqs1}), (\ref{KGfeqs2})
to obtain two independent second-order differential equations for the combinations
$\Psi_{1}+\Psi_{2}$ and $\Psi_{1}-\Psi_{2}$, which look like they should not lead
to any pathologies.

Relation (\ref{equivcond}) seems to be rather unphysical (indeed,
a consistent localization of the zero modes of fermion fields
demands (\ref{cond1}), see also, for example,
\cite{DRT,CG,Andrianov:2003hx}; while it is unclear what could be
the physical motivation for the condition $F_{1}(z)\equiv
F_{2}(z)\not\equiv 0$); except for the case $F_{1}(z)\equiv
F_{2}(z)\equiv 0$, which means that all the fermion fields can
freely propagate in the bulk. Thus, to examine this case in more
detail let us simplify the task and take $\sigma(z)\equiv 0$,
$F_{1}(z)\equiv F_{2}(z)\equiv 0$ and
$v(z)\not\equiv\textrm{const}$ (for the case $\sigma(z)\equiv 0$ the
latter condition does not satisfy (\ref{cond2})). The
corresponding equations of motion, following from
(\ref{KGfeqs1}) and (\ref{KGfeqs2}), take the form
\begin{eqnarray}\label{KGfeqs1flat2}
-\Box(\Psi_{1}+\Psi_{2})+\partial_{5}^{2}(\Psi_{1}+\Psi_{2})-h^2v^{2}(\Psi_{1}+\Psi_{2})+hv'\gamma^{5}(\Psi_{1}+\Psi_{2})=0,\\ \label{KGfeqs2flat2}
-\Box(\Psi_{1}-\Psi_{2})+\partial_{5}^{2}(\Psi_{1}-\Psi_{2})-h^2v^{2}(\Psi_{1}-\Psi_{2})-hv'\gamma^{5}(\Psi_{1}-\Psi_{2})=0.
\end{eqnarray}
Using these equations it is not difficult to show that,  according
to the orbifold symmetry conditions (\ref{sym1}), (\ref{sym2}),
the fields $\Psi_{1}$ and $\Psi_{2}$ can be decomposed into the
Kaluza-Klein modes as
\begin{eqnarray}\label{KKdecf1}
\Psi_{1}=\sum_{n}\left(f_{+}^{n}(z)\psi_{L}^{n}(x)-f_{-}^{n}(z)\psi_{R}^{n}(x)\right),\\ \label{KKdecf2}
\Psi_{2}=\sum_{n}\left(f_{-}^{n}(z)\psi_{L}^{n}(x)+f_{+}^{n}(z)\psi_{R}^{n}(x)\right),
\end{eqnarray}
where $\gamma^{5}\psi_{L}=\psi_{L}$, $\gamma^{5}\psi_{R}=-\psi_{R}$,
\begin{equation}
f_{+}^{n}(z)=f^{n}(z)+f^{n}(-z),\qquad f_{-}^{n}(z)=f^{n}(z)-f^{n}(-z)
\end{equation}
and the function $f^{n}(z)$ is a periodic continuously differentiable solution to the equation
\begin{equation}\label{diffeqf}
m_{n}^{2}f^{n}+\partial_{5}^{2}f^{n}-h^2v^{2}f^{n}+hv'f^{n}=0
\end{equation}
in the interval $[-L,L]$, corresponding to the eigenvalue $m_{n}^2$ (recall that
$v(-z)=v(z)$). According to the  general theory \cite{CL}, the functions  $f^{n}(z)$ make up an orthonormal set of
eigenfunctions  for equation (\ref{diffeqf}), the lowest
eigenvalue $m_{0}$ being simple (which means that we get only one
fermion with mass $m_{0}$ in the effective four-dimensional
theory). Moreover, it is not difficult to show that $m_{0}^2>0$
for (\ref{diffeqf}). Thus, the corresponding free theory seems to have
no obvious pathologies. But the chiral structure of the zero modes of
the fields $\Psi_{1}$ and $\Psi_{2}$ in (\ref{KKdecf1}), (\ref{KKdecf2})
differs from that in (\ref{massf1}), (\ref{massf2}). This difference
leads to certain problems when taking into account the interactions
with the gauge fields. This issue will be discussed in the next section.

What can happen when the decoupling of the equations of motion for
the components of the fermion fields seems to be impossible, at
least in the standard way (like the one presented in Appendix~B)?
To show it, let us again simplify the task and consider the case
$\sigma(z)\equiv 0$ (the flat five-dimensional space-time). In this
case equations (\ref{KGfeqs1}) and (\ref{KGfeqs2}) take the form
\begin{eqnarray}\label{KGfeqs1flat}
-\Box\Psi_{1}+\partial_{5}^{2}\Psi_{1}+F_{1}'\gamma^{5}\Psi_{1}-\left(F_{1}^{2}+h^2v^{2}\right)\Psi_{1}+hv'\gamma^{5}\Psi_{2}-hv\left(F_{1}+F_{2}\right)\Psi_{2}=0,\\ \label{KGfeqs2flat}
-\Box\Psi_{2}+\partial_{5}^{2}\Psi_{2}+F_{2}'\gamma^{5}\Psi_{2}-\left(F_{2}^{2}+h^2v^{2}\right)\Psi_{2}+hv'\gamma^{5}\Psi_{1}-hv\left(F_{1}+F_{2}\right)\Psi_{1}=0.
\end{eqnarray}
The equation for, say, the first component of the field
$\Psi_{1}$ can be easily obtained and turns out to be the fourth-order differential equation
\begin{eqnarray}\label{hdereq}
\left[\Box-\partial_{5}^{2}-F_{2}'+F_{2}^{2}+h^2v^{2}\right]\frac{1}{hv(F_{1}+F_{2})-hv'}
\left[\Box-\partial_{5}^{2}-F_{1}'+F_{1}^{2}+h^2v^{2}\right]\Psi_{1}^{(1)}\\ \nonumber-\left(hv(F_{1}+F_{2})-hv'\right)\Psi_{1}^{(1)}=0.
\end{eqnarray}
Analogous equations can be obtained for the other components of
the field $\Psi_{1}$ and for the components of the field
$\Psi_{2}$ (though, as in the four-dimensional case, not all the components of the fields $\Psi_{1}$ and $\Psi_{2}$ are completely independent, some of them can be expressed through the other: for example, the field $\Psi_{2}$ can be restored from the field $\Psi_{1}$ with the help of initial equation (\ref{feqs1})).  It is obvious that even in the flat case
$\sigma(z)\equiv 0$ the form of equation (\ref{hdereq}) poses a
question about the possibility of a mathematically consistent
isolation of the physical degrees of freedom of the theory.
Moreover, since in many cases higher-derivative theories contain
pathologies (such as ghosts; see, for example,
\cite{Woodard:2006nt} for details), one may expect pathological
behavior in the case under consideration too. On the other hand, usually the systems described by fourth-order equations of motion like the one in (\ref{hdereq}) have more degrees of freedom than the systems described by second-order equations of motion like the one in (\ref{KGfeqs1}) with $v(z)\equiv 0$, so in principle one can expect an increase of the number even of the zero modes in addition to other possible pathologies (i.e., one could imagine that there would appear, say, two electrons in the effective four-dimensional theory). We have failed to solve such fourth-order equations of motion analytically even in the simplest cases and we have not an explicit example which could support this statement, however, such a possibility is not excluded by the general reasonings. Thus, in our
opinion, one should avoid the appearance of such fourth-order
differential equations when constructing multidimensional models, at least
to be sure that the resulting theory is devoid of any pathologies
and the physical degrees of freedom can be isolated in a
mathematically consistent way using the well-developed theory of
second-order differential equations.

An important comment is in order here. In many brane world models
the expansion in the Kaluza-Klein modes for the fields $\Psi_{1}$
and $\Psi_{2}$ is performed without taking into account the
interaction with the Higgs field (indeed, in the case $v(z)\equiv
0$ the corresponding differential equations are indeed
second-order and one does not expect any pathologies). When the interaction with the Higgs field is taken back into
account, all the off-diagonal entries of the corresponding
(infinite) mass matrix are, in the general case, nonzero, which is
not an unexpected result, because the orthogonality conditions for
the case $v(z)\equiv 0$ are not valid for the case $v(z)\not\equiv
0$, $v(z)\not\equiv\tilde v e^{-\sigma}$ (including the
generalized functions like the delta-function). It is clear that, in principle, the resulting
fields do not represent the physical degrees of freedom of the
theory and cannot be used for consistent calculations, which poses the question about
the diagonalization of the mass matrix in the effective four-dimensional theory by algebraic (maybe perturbative) methods.
However, as it was noted above, in the general case the systems described by fourth-order equations of motion like the one in (\ref{hdereq})
have more degrees of freedom than the systems described by second-order equations of motion like the one in (\ref{KGfeqs1}) with $v(z)\equiv 0$. The latter makes the use of the perturbation theory and the subsequent diagonalization of the mass matrix questionable, because in the general case this mass matrix, obtained using the solutions of equations (\ref{KGfeqs1}) and (\ref{KGfeqs2}) with $v(z)\equiv 0$, may not include all the degrees of freedom (including possible pathological modes) described by equations (\ref{KGfeqs1}) and (\ref{KGfeqs2}) with $v(z)\not\equiv 0$. We note that it is a nonperturbative effect, which may appear no matter
what is the relation between the energy scale of the Higgs field
and the typical energy scale of five-dimensional theory.\footnote{There is a simple algebraic example, demonstrating an analogous nonperturbative effect. The equation $x-1=0$ has only one real root, whereas the equation $\alpha x^2+x-1=0$ with $|\alpha|\ll 1$ has two real roots. The first root $x\approx 1-\alpha$ indeed can be obtained perturbatively, whereas it is not so for the other root $x\approx-\frac{1}{\alpha}-1+\alpha$.} In some
sense this situation is similar to the case of $U(1)$ massless
gauge field: if one adds the mass term to the action, the
third degree of freedom arises, no matter how small the mass of
such a ``photon'' is. It is impossible to trace such an effect by first performing the KK decomposition in the vacuum with $v(z)\equiv 0$ and only then considering the interaction with the nonzero vacuum solution of the Higgs field. For this reason we think that the only
consistent way of deriving an effective four-dimensional action in
brane world models is to consider first the vacuum solution for
the Higgs field (and for other fields with nonzero vacuum
solutions, if they exist) and only then to perform the
Kaluza-Klein decomposition (if it is possible) checking the
absence of pathologies at least in the free theory.

\section{Interactions in the effective theory}
To demonstrate in a simple way, how possible latent problems
can pop up in the four-di\-mensional effective theory, corresponding to equations (\ref{KGfeqs1flat2}) and (\ref{KGfeqs2flat2}), let us
consider a five-dimensional action, describing fermion fields
minimally coupled to the $SU(2)\times U(1)$ gauge fields in the flat ($\sigma(z)\equiv
0$) space-time:
\begin{eqnarray}\label{action2}
S=\int
d^{4}xdz\left(i{\bar{\hat\Psi}_{1}}\Gamma^{M}D_{M}{\hat\Psi}_{1}+
i\bar\Psi_{2}\Gamma^{M}D_{M}\Psi_{2}-\sqrt{2}h\left[\left({\bar{\hat\Psi}_{1}}
H\right)\Psi_2+\textrm{h.c.}\right]\right).
\end{eqnarray}
Here the $SU(2)$ doublet, constructed from five-dimensional spinors, is denoted by
\begin{eqnarray}
{\hat\Psi}_{1}=\left(
\begin{array}{l}
\Psi_{1}^{\nu} \\ \Psi_{1}^{\psi}\\
\end{array}
\right),
\quad {\bar{\hat\Psi}_{1}}=\left(\bar\Psi_{1}^{\nu}\,,\,\bar
\Psi_{1}^{\psi}\right)
\end{eqnarray}
and the five-dimensional $SU(2)$ singlet is denoted by $\Psi_{2}$. The covariant derivatives are
defined by
\begin{eqnarray}
D_{M}{\hat\Psi}_{1}&=&\left(\partial_{M}-ig\frac{\tau^{a}}{2}A_{M}^{a}+i\frac{g'}{2}B_{M}\right){\hat\Psi}_{1},\\
D_{M}\Psi_{2}&=&\left(\partial_{M}+ig'B_{M}\right)\Psi_{2}.
\end{eqnarray}
The vacuum solution for the Higgs field is supposed to have the form
\begin{equation}\label{vev2}
H_{0}\equiv\left(0\atop \frac{v(z)}{\sqrt{2}}\right)
\end{equation}
with $v(z)\not\equiv\textrm{const}$.

First, let us consider the free theory. We will be interested only
in the lowest mode sector, so we neglect all the higher Kaluza-Klein modes of gauge and fermion fields. According
to (\ref{KGfeqs1flat2})--(\ref{diffeqf}), we can represent the fermion zero
modes as (below we will omit the superscript ``0'' for
the zero ($n=0$) modes of the fields)
\begin{eqnarray}\label{substferm}
{\hat\Psi}_{1}(x,z)=\left(
\begin{array}{l}
\xi\nu_{L}(x) \\ f_{+}(z)\psi_{L}(x)-f_{-}(z)\psi_{R}(x)\\
\end{array}\right),\quad \Psi_{2}(x,z)=f_{-}(z)\psi_{L}(x)+f_{+}(z)\psi_{R}(x),
\end{eqnarray}
where $\nu_{L}=\gamma^{5}\nu_{L}$. The factor
$\xi=\frac{1}{\sqrt{2L}}$  is just the canonically normalized wave
function of the zero mode of the field $\Psi_{1}^{\nu}$ (since
this field does not interact with the vacuum solution of the Higgs
field $H$, in the flat background and with $F_{1}(z)\equiv
F_{2}(z)\equiv 0$ its wave function is a constant). Substituting (\ref{substferm}) into (\ref{action2}), dropping the
terms with gauge fields and integrating over the coordinate of the
extra dimension, one can obtain the standard
four-dimensional action for the free fermion fields\footnote{In
the derivation of action (\ref{action2-4d}) it is convenient to
use the equations $f_{+}'+hvf_{-}=-mf_{-}$,\\
$f_{-}'+hvf_{+}=mf_{+}$, which are fulfilled if equation
(\ref{diffeqf}) holds.}
\begin{eqnarray}\label{action2-4d}
S=\int d^{4}x\left(i\bar{\nu_{L}}\gamma^{\mu}\partial_{\mu}\nu_{L}+i\bar{\psi}\gamma^{\mu}\partial_{\mu}\psi-m\bar{\psi}\psi\right),
\end{eqnarray}
where $\psi=\psi_{L}+\psi_{R}$. In order to have the canonically
normalized kinetic term of the field $\psi$ in (\ref{action2-4d}),
the condition
\begin{equation}\label{norm}
a^{2}+b^{2}=1,\qquad a^{2}=\int dz f_{+}^{2}(z),\qquad b^{2}=\int dz f_{-}^{2}(z).
\end{equation}
must be fulfilled. The mass $m$ is the eigenvalue of the problem
(\ref{diffeqf}), corresponding to the eigenfunction $f(z)$. By
tuning the coupling constant $h$ one can, in principle, get the
desired value of the mass $m$.

Now let us turn to examining the interactions with the gauge bosons.
In order to isolate the effects caused only by the fermions, below we choose the following
ansatz for the zero modes of the gauge fields:
\begin{eqnarray}\label{substgb}
A^{a}_{\mu}(x,z)\equiv A^{a}_{\mu}(x),\quad B_{\mu}(x,z)\equiv B_{\mu}(x);
\end{eqnarray}
for simplicity we will also drop the components $A^{a}_{5}(x,z)$ and $B_{5}(x,z)$ of
these fields. We do not discuss
here possible ways for obtaining this ansatz in a consistent way.
For example, one may simply imagine that there exists a second
Higgs field, interacting with the gauge fields only, which
provides the necessary forms of the corresponding wave functions.

Passing to the physical degrees of freedom (\ref{gaugephys}),
using (\ref{substferm}), (\ref{substgb}) and integrating over the
coordinate of the extra dimension, we can obtain the effective
four-dimensional action, describing the interaction of the zero
mode fermions with the gauge bosons. We do not present the explicit
calculations here, they are straightforward. It is not difficult
to show that the electromagnetic coupling constant appears to be
the same as in the SM (from here and below ``the same as in the
SM'' means that it can be expressed through the constants $g$ and
$g'$ exactly in the same way as it happens in the SM). The
coupling constant of the interaction with the charged gauge bosons is
found to be
\begin{equation}
g\frac{1}{\sqrt{2L}}\int dz f_{+}(z)
\end{equation}
instead of $g$ in the SM (recall that the wave function,
corresponding  to the field $\nu_{L}$, is just
$\frac{1}{\sqrt{2L}}$). As for the interaction of the field $\psi$ with the neutral
gauge boson $Z$, the vector coupling constant appears to be the
same as in the SM, whereas the axial coupling constant has the
form
\begin{equation}
g_{A}=g^{SM}_{A}(a^{2}-b^{2}).
\end{equation}

It is clear that in the general case $c=\frac{1}{\sqrt{2L}}\int dz
f_{+}(z)\ne 1$  and, according to (\ref{norm}), $a^{2}-b^{2}\ne
1$. Meanwhile, in order to get rid of the difference with the well
known parameters of the SM, one should have $c=1$ and
$a^{2}-b^{2}=1$ (or the values which are close to unity with a
good accuracy). The latter can be achieved if $b=0$, which means
that $f(z)\equiv\textrm{const}$; in this case $c=1$ too. But, according
to (\ref{diffeqf}), the condition $f(z)\equiv\textrm{const}$ means that
$v(z)\equiv\textrm{const}$, which again corresponds to (\ref{cond2})
with $\sigma(z)\equiv 0$. Thus, the farther $v(z)$ from a constant is, the
farther the values of the corresponding coupling constants are
from those of the SM.

It is obvious that for the non-flat case $\sigma(z)\not\equiv 0$ with
$\partial_{5}(e^{\sigma}v(z))\not\equiv 0$ the problems,
completely analogous to those described above, are also expected
to arise in the four-dimensional effective theory, though the
calculations appear to be much more bulky than in the case
$\sigma(z)\equiv 0$. The only difference in the warped case is that
now the farther $v(z)$ from $\sim e^{-\sigma}$ is, the farther the
values of the corresponding coupling constants are from those of
the SM\footnote{This reasoning is valid only for the case
$F_{1}(z)\equiv F_{2}(z)\equiv 0$. If $F_{1}(z)\equiv
F_{2}(z)\not\equiv 0$, there should be deviations of the coupling
constants from those of the SM even for
$\partial_{5}(e^{\sigma}v(z))\equiv 0$ due to the nonzero rotation
angle $\theta$, see Appendix~B.} (note that in a realistic theory
with $\partial_{5}(e^{\sigma}v(z))\not\equiv 0$ the shapes of the
zero mode gauge boson wave functions also differ from a constant,
which provides additional deviations from the standard values of
the SM parameters like in \cite{Csaki:2002gy,Burdman:2002gr}). For
example, the natural choice $v(z)\equiv\textrm{const}$ may lead, in
principle, either to an unacceptable theory or put it out of the
reach of the present day experiments.

As a final remark to this section, let us draw an analogy
with  the case of gauge fields. As it was mentioned in Section~2,
the modification of the shapes of the gauge boson wave functions
due to the interaction with the vacuum solution of the Higgs field
may affect the four-dimensional effective theory considerably
\cite{Csaki:2002gy,Burdman:2002gr}. In the most cases the interaction of fermion
fields with the vacuum solution of the Higgs field affects the
corresponding effective theory either in an analogous way (by a
modification of the zero mode wave functions and  the chiral
structure), or even more dramatically, leading to pathologies like
those discussed in Section~3. The latter clearly indicates that the interaction of fermion fields with the vacuum
solution of the Higgs field should be treated much more carefully
than it is usually done.

\section{Conclusion and final remarks}
As it was demonstrated in the previous sections, the only obvious
possibility to automatically get a self-consistent, from the
theoretical point of view, four-dimensional SM in a
five-dimensional brane world model (i.e., without possible
pathologies in the free theory and with the correct couplings) is
to have a vacuum solution of form (\ref{higgsprof}) for the Higgs
field together with (\ref{cond1}). Condition (\ref{cond1})
corresponds, in fact, both to the case of localized fermion zero
modes and to the case when the fermion fields can freely propagate
in the bulk (if $F_{1}(z)\equiv F_{2}(z)\equiv 0$).\footnote{As it
was shown in Appendix~B, there may exist other exceptions in
equations (\ref{KGfeqs1}), (\ref{KGfeqs2}), leading to
second-order differential equations of motion. But, according to
the results presented above, it is improbable that such unnatural
cases could lead to a completely acceptable effective theory.} It
is not difficult to check that in this case the zero mode fermion
and gauge boson sectors of the resulting effective theory indeed
exactly reproduce those of the SM, including the interactions, at
least for the case of the standard form of five-dimensional gauge
invariant action (of course, if the localizing functions $F(z)$
are one and the same for all the fields, corresponding to
different four-dimensional SM particles). The corresponding
calculations are straightforward and we do not present them here.
The self-consistency and automatic conformity to the SM of the
zero mode fermion and gauge boson sectors of the resulting
four-dimensional effective theory for (\ref{cond1}) and
(\ref{cond2}), follow from the fact that only in this case
equations (\ref{KGfeqs1}) and (\ref{KGfeqs2}) for the components
of the five-dimensional fermions $\Psi_{1}$ and $\Psi_{2}$
decouple, which provides the correct chiral structure of the
corresponding lowest Kaluza-Klein modes. Deviations from these
conditions may lead either to pathologies or to a variance between
the resulting zero mode four-dimensional effective theory and the
SM. The latter may result in severe constraints on the parameters
of five-dimensional theory and put it, in principle, out of the
reach of the present day experiments in full analogy with how in
happens in the gauge boson sector
\cite{Csaki:2002gy,Burdman:2002gr}.

It should be noted that the exponential profile of  the vacuum
solution for the Higgs field, leading to
its localization near the TeV brane in the Randall-Sundrum model
\cite{Randall:1999ee}, was discussed earlier \cite{Huber:2000ie}.
Our results demonstrate that in order to have a possibility to
localize the zero modes of fermion fields in a consistent way
(using an appropriate form of the function $F(z)$ in
(\ref{KGfeqs1a}) and (\ref{KGfeqs2a})), the profile of the vacuum
solution of the Higgs field should have exactly the form
(\ref{higgsprof}).

Unfortunately, the restriction on the vacuum profile of the Higgs
field poses several problems for the case of localized fermion zero modes (i.e., when $F(z)\not\equiv 0$), which should be necessarily
addressed. First, the profile of the vacuum solution for the Higgs
field appears  to be strongly related to the form of the warp
factor of the model. This demands an extra fine-tuning for the
scalar field potential. Indeed, even for the simplest case of the
Randall-Sundrum setup \cite{Randall:1999ee} with $\sigma(z)=-k|z|$
and without taking into account the backreaction of the Higgs
field on the background metric, the fine-tuned bulk Higgs
potential should have the form
$$
V(H^{\dag}H)=-3k^2H^{\dag}H
$$
to get the vacuum solution (\ref{higgsprof}). Of course, one should also add fine-tuned
brane-localized potentials, including a term specifying the
value of the constant $\tilde v$ (at least on one of the branes).
If one takes more realistic cases of stabilized models, in which
the warp factors have a more complicated form (like the one in
\cite{DeWolfe:1999cp}), the form of the Higgs scalar field potential appears
to be such that it can not be represented in an analytical form.
Of course, such a situation looks unnatural, at least in the
absence of a symmetry that can support such a fine-tuning of
the Higgs potential. Moreover, backreaction of the Higgs field affects the background metric, whereas quantum corrections modify the scalar field potential and, consequently, the vacuum solution for the Higgs field. Both effects lead to breakdown of the fine-tuned relation between the warp factor and the vacuum solution for the Higgs field.

Second, in a scenario with localized zero modes of fermion fields
the Higgs field and the stabilizing scalar field cannot be
unified, as it was proposed in \cite{Geller:2013cfa}. Indeed, a
consistent stabilization mechanism (like the one proposed in
\cite{DeWolfe:1999cp}) is based on fixing the values of the
stabilizing scalar fields on the branes (at the points $z=0$ and
$z=L$). On the other hand, warped brane world models are
interesting if the function $e^{2\sigma}$ has exponentially
different values on the branes. The latter means that by taking
the Higgs field as the stabilizing field in such a theory, one
introduces a new hierarchy into the model (because $v(0)\ll
v(L)$). For example, the Randall-Sundrum model
\cite{Randall:1999ee} was proposed to solve the hierarchy problem
of gravitational interaction, so it also looks unnatural to add an
extra hierarchy into such a model. Moreover, in order to get a
massive radion, one should take into account the backreaction of
the stabilizing field on the background metric
\cite{DeWolfe:1999cp}. But if the stabilizing field is the Higgs
field with vacuum solution (\ref{higgsprof}), then the range of
allowed scalar field potentials and warp factors narrows
considerably (which clearly follows from the self-consistent
system of equations for the background configuration of the metric
and the stabilizing scalar field, which can be found in
\cite{DeWolfe:1999cp}).

One may suppose that if at least the fermion fields are located
exactly on the brane, then the Higgs field can also be located on
the brane, which looks as a solution to the problem (of course,
if we do not take into account the gauge fields, see
\cite{Csaki:2002gy}). However, the only realistic
field-theoretical mechanism of fermion localization, which can be
used for calculations, is based on the idea that initially the
fermion fields propagate in the whole five-dimensional space-time,
whereas only the lowest modes appear to be localized on the brane
due to an interaction with some defect (for example, with a domain
wall like in \cite{RS}), see the discussion in Section~3. This is
exactly the situation, which is realized in equations
(\ref{massf1}),~(\ref{massf2}), so one can take an appropriate
form of the function $F(z)$ to make the width of the wave function
of the localized fermion as small as necessary. Meanwhile, the
profile of the Higgs field does not depend on the form of the
function $F(z)$, so even for an extremely narrow wave function of
a localized mode (which taken squared can be even approximated by
the delta-function for calculations) the ``right'' profile of the
Higgs field, which does not lead to fourth-order differential
equations, should still have the form (\ref{higgsprof}). The
latter poses a question whether there exists a field-theoretical
mechanism of fermion localization, leaving more freedom for the
choice of a vacuum profile of the Higgs field in the extra
dimension.

It should be noted that the only obvious exception is the model
with the flat five-dimensional background metric like the one
proposed in \cite{Appelquist:2000nn}, for which the
four-dimensional SM can be constructed from a five-dimensional theory without unnatural fine-tunings and restrictions. In
such a case the vacuum solution for the Higgs field must be just a
constant, which admits a variety of scalar field potentials
including the standard Higgs potential (but leaving unsolved the
problem of the  stabilization of the extra dimension  size).

\section*{Acknowledgements}
The authors are grateful to E.~Boos and S.~Keizerov for
discussions   and useful remarks. The work was supported by grant
14-12-00363 of Russian Science Foundation.

\section*{Appendix A: Mass of the lowest localized fermion mode}
Let us take equation (\ref{KGfeqs1a}) and substitute
$\Psi_{1}(x,z)=\psi_{L}(x)f_{L}(z)$ with
$\Box\psi_{L}+\mu^2\psi_{L}=0$ and $\gamma^{5}\psi_{L}=\psi_{L}$
into it. We get
\begin{equation}
(\mu^2-{\tilde h}^2{\tilde
v}^{2})f_{L}+e^{\sigma}(\partial_{5}+2\sigma')e^{\sigma}(\partial_{5}+2\sigma')f_{L}+e^{\sigma}\partial_{5}(e^{\sigma}F)f_{L}
-e^{2\sigma}F^{2}f_{L}=0.
\end{equation}
Multiplying this equation by $e^{3\sigma}f_{L}$, integrating over the
coordinate of the extra dimension $z$ and performing integration
by parts in two terms, we arrive to the following equality:
\begin{equation}
(\mu^2-{\tilde h}^2{\tilde v}^{2})\int dz
e^{3\sigma}f_{L}^{2}=\int dz
e^{5\sigma}\left(f_{L}'+2\sigma'f_{L}+Ff_{L}\right)^{2}.
\end{equation}
Since both integrals are nonnegative, we get $\mu^2-{\tilde
h}^2{\tilde v}^{2}\ge 0$,  which means that the lowest mode indeed
has mass ${\tilde h}\tilde v$. A completely analogous procedure
can be performed for the other substitution
$\Psi_{1}(x,z)=\psi_{R}(x)f_{R}(z)$ with
$\Box\psi_{R}+\mu^2\psi_{R}=0$ and $\gamma^{5}\psi_{R}=-\psi_{R}$,
as well as for the field $\Psi_{2}$.

\section*{Appendix B: Decoupling of the equations of motion for fermions}
Let us consider equations (\ref{KGfeqs1}) and (\ref{KGfeqs2}) for
the left-handed fermions such that
$\Psi_{1}^{L}=\gamma^{5}\Psi_{1}^{L}$,
$\Psi_{2}^{L}=\gamma^{5}\Psi_{2}^{L}$. These equations can be
rewritten in the matrix form as
\begin{equation}\label{eqsleft}
{\hat L}(x,z)\left(\Psi_{1}^{L}\atop \Psi_{2}^{L}\right)+{\hat M}(z)\left(\Psi_{1}^{L}\atop \Psi_{2}^{L}\right)=0,
\end{equation}
where ${\hat L}(x,z)$ is a diagonal operator with equal diagonal
elements, which includes derivatives, and the matrix ${\hat M}(z)$
looks like
\begin{equation}
{\hat M}(z) =\begin{pmatrix}
q_{1}(z)& p(z)\\
p(z)&q_{2}(z)\\
\end{pmatrix}
\end{equation}
with
\begin{eqnarray}
q_{1}(z)&=&e^{\sigma}\partial_{5}(e^{\sigma}F_{1}(z))-e^{2\sigma}F_{1}^{2}(z),\\
q_{2}(z)&=&e^{\sigma}\partial_{5}(e^{\sigma}F_{2}(z))-e^{2\sigma}F_{2}^{2}(z),\\
p(z)&=&he^{\sigma}\partial_{5}(e^{\sigma}v(z))-he^{2\sigma}v(z)\left(F_{1}(z)+F_{2}(z)\right).
\end{eqnarray}
The form of equation (\ref{eqsleft}) suggests that the decoupling
of the equations of motion for the fermion fields is equivalent to
the diagonalization of the matrix ${\hat M}(z)$, so the question
is how to diagonalize the matrix ${\hat M}(z)$ except for the
obvious case $p(z)\equiv 0$. Since this matrix is symmetric, it
can be diagonalized with the help of a rotation matrix $\hat R$
such that
\begin{equation}
{\hat R}^T{\hat M}{\hat R}=\textrm{diag}(\lambda_{1},\lambda_{2}),\qquad {\hat R}=\begin{pmatrix}
\cos \theta& -\sin \theta\\
\sin \theta&\cos \theta\\
\end{pmatrix}.
\end{equation}
The eigenvalues of the matrix ${\hat M}(z)$ can be easily found by the standard procedure and take the form
\begin{eqnarray}
\lambda_{1,2}(z)&=&\frac{q_{1}+q_{2}\pm\sqrt{(q_{1}-q_{2})^2+4p^{2}}}{2}.
\end{eqnarray}
A very important point is that the rotation angle $\theta$ should
not depend on the coordinate of the extra dimension $z$, otherwise
the rotation matrix $\hat R$ would not pass through the operator
$\hat L$, which contains derivatives. Since it is the
eigenvectors of the matrix ${\hat M}$ that form the
rotation matrix $\hat R$, it is necessary to find conditions under
which these eigenvectors do not depend on the coordinate of the
extra dimension. From the equation, say, for the first eigenvalue
and eigenvector
\begin{equation}
{\hat M}(z)\left(\cos\theta\atop \sin\theta\right)=\lambda_{1}(z)\left(\cos\theta\atop \sin\theta\right)
\end{equation}
we can easily get
\begin{equation}
\tan\theta=\frac{q_{2}-q_{1}+\sqrt{(q_{1}-q_{2})^2+4p^{2}}}{2p}.
\end{equation}
It is clear that in the general case the angle $\theta$ depends on
the coordinate of the extra dimension. An obvious exception is in
general $q_{1}(z)\equiv q_{2}(z)$.

Thus, according to the results, presented above, in the general
case we have two conditions for the left-handed fermions, for
which the mixing matrix ${\hat M}$ is either diagonal or can be
diagonalized in the standard way. They are
\begin{eqnarray}\label{fermc1}
\partial_{5}(e^{\sigma}v)&=&e^{\sigma}v\left(F_{1}+F_{2}\right),\\ \label{fermc2}
\partial_{5}(e^{\sigma}(F_{1}-F_{2}))&=&e^{\sigma}(F_{1}^{2}-F_{2}^{2}).
\end{eqnarray}

The procedure, completely analogous to the one presenter above,
can be also performed for the right-handed fermions
($\Psi_{1}^{R}=-\gamma^{5}\Psi_{1}^{R}$,
$\Psi_{2}^{R}=-\gamma^{5}\Psi_{2}^{R}$), leading to
\begin{eqnarray}\label{fermc3}
\partial_{5}(e^{\sigma}v)&=&-e^{\sigma}v\left(F_{1}+F_{2}\right),\\ \label{fermc4}
\partial_{5}(e^{\sigma}(F_{1}-F_{2}))&=&-e^{\sigma}(F_{1}^{2}-F_{2}^{2}).
\end{eqnarray}
The most general and simple model-independent conditions,
following from (\ref{fermc1})--(\ref{fermc4}), are just
(\ref{cond1}), (\ref{cond2}) or (\ref{equivcond}), the latter case
giving $\theta=\pm\frac{\pi}{4}$, which corresponds to the
combinations of the fields in (\ref{KGfeqs1flat2}),
(\ref{KGfeqs2flat2}). Of course, in principle one may consider
other combinations of conditions (\ref{fermc1})--(\ref{fermc4}):
(\ref{fermc1}) together with (\ref{fermc4}) or (\ref{fermc2})
together with (\ref{fermc3}); or there may exist other specific
choices of the functions $F_{1}(z)$, $F_{2}(z)$ and $v(z)$ leaving
the rotation angle $\theta$ independent on the coordinate of the
extra dimension. But such cases seem to be much more unnatural,
while they also lead to relations between the functions
$F_{1}(z)$, $F_{1}(z)$, $v(z)$ and $\sigma(z)$.

\end{document}